\documentclass[reqno]{amsart}
\usepackage{amssymb}
\usepackage{verbatim}

\newcommand{\bea}{\begin{eqnarray}}
\newcommand{\eea}{\end{eqnarray}}
\newcommand{\beann}{\begin{eqnarray*}}
\newcommand{\eeann}{\end{eqnarray*}}
\newcommand{\BT}[1]{\begin{tabular}{#1}}

\newcommand{\ET}{\end{tabular}}
\newcommand{\bdm}{\begin{displaymath}}
\newcommand{\edm}{\end{displaymath}}

\newcommand{\q}{\quad}

\begin{document}

\title{Automorphisms of the fine grading of $sl(n,\mathbb{C})$
        associated with the generalized Pauli matrices}

\author{Miloslav Havl\'{\i}\v{c}ek}
\address{Department of Mathematics and Doppler Institute,
         Faculty of Nuclear Sciences and Physical
         Engineering,  Czech Technical University,
         Trojanova 13, 120 00  Praha 2,
         Czech Republic}
\email{havlicek@km1.fjfi.cvut.cz}

\author{Ji\v{r}\'\i\  Patera}
\address{Centre de Recherches Math\'ematiques,
         Universit\'e de Montr\'eal,
         C.P.6128-Centre ville,
         Montr\'eal, H3C\,3J7, Qu\'ebec, Canada}
\email{patera@crm.umontreal.ca}

\author{Edita Pelantov\'a}
\address{Department of Mathematics and Doppler Institute,
         Faculty of Nuclear Sciences and Physical
         Engineering,  Czech Technical University,
         Trojanova 13, 120 00  Praha 2,
         Czech Republic}
\email{pelantova@km1.fjfi.cvut.cz}

\author{Ji\v{r}\'{\i} Tolar}
\address{Department of Physics and Doppler Institute,
         Faculty of Nuclear Sciences and Physical
         Engineering,  Czech Technical University,
         B\v{r}ehova 7, 115 19  Praha 1,
         Czech Republic}
\email{tolar@br.fjfi.cvut.cz}

\date{\today}

\begin{abstract}
We consider the grading of $sl(n,\mathbb{C})$ by the group $\Pi_n$
of generalized Pauli matrices. The grading decomposes the Lie
algebra into $n^2-1$ one--dimensional subspaces. In the article we
demonstrate that the normalizer of grading decomposition of
$sl(n,\mathbb{C})$ in $\Pi_n$ is the group $SL(2, \mathbb{Z}_n)$,
where $\mathbb{Z}_n$ is the cyclic group of order $n$.

As an example we consider $sl(3,\mathbb{C})$ graded by $\Pi_3$ and
all contractions preserving that grading. We show that the set of
48 quadratic equations for grading parameters splits into just two
orbits of the normalizer of the grading in $\Pi_3$.
\end{abstract}

\maketitle

\section{Introduction}
Among the gradings of reductive Lie algebras over the complex
number field and the simultaneous gradings of their representation
spaces, by far the most important ones are the gradings by maximal
torus. In the case of the Lie algebra it is also called root or
Cartan decomposition. Such a grading means a decomposition into
eigenspaces of the maximal torus. For a greater part of the past
century such gradings have been the workhorses of the theory and
applications.

Typical role a Lie algebra plays in physics is the algebra of
infinitesimal symmetries of a physical system, which themselves are
described in terms elements of representation spaces, eigenvectors
of the maximal torus. The corresponding eigenvalues are then the
quantum numbers.

The question about existence of other gradings, like those by
maximal torus (called fine gradings), has been raised
systematically in \cite{PZ1} and solved for the simple Lie
algebras of over $\mathbb{C}$ in \cite{HPP1,HPP2,HPP3}
and recently also for
the real number field in \cite{HPP4,HPP6}.

Gradings of Lie  algebras are closely related with their
automorphisms. In a seminal paper \cite{PZ1} in 1989, it was shown
that the finest gradings (called fine) of finite-dimensional
simple Lie algebras  ${\mathcal L}$ can be classified  (up to
equivalence generated by elements of Aut ${\mathcal L}$) by the
Maximal Abelian groups of Diagonable automorphisms of ${\mathcal
L}$, briefly  the MAD--groups. In general, the MAD--groups are
composed, besides subgroups of the maximal torus, by well--known
outer automorphisms \cite{H}, and by elements of finite order
(EFO's) in the corresponding Lie groups. Since the conjugacy
classes of EFO's were systematically described in \cite{K} (see
also \cite{H}), it was not difficult to classify the fine gradings
in the lowest cases like $sl(2,\mathbb{C})$ and $sl(3,\mathbb{C})$
\cite{PZ2,4,W1}.

A prominent role in the grading problem of simple Lie algebras is
played by the finite group $\Pi_n$ of $n^3$ matrices
$\mathbb{C}^{n\times n}$. A subset of $\Pi_n$, consisting of
$n^2-1$ traceless matrices, can be taken as a basis of
$sl(n,\mathbb{C})$. Since these traceless matrices of
are used as a basis of an Lie algebra, they can be
normalized in any convenient way. In particular, for $n=2$
the Pauli matrices of $sl(2,\mathbb{C})$ are obtained.

Most importantly, the adjoint action of $\Pi_n$, though
non--Abelian in general, induces an Abelian action on
$sl(n,\mathbb{C})$. The fine grading of $sl(n,\mathbb{C})$, which
arises in this way, decomposes the algebra into one--dimensional
subspaces generated by  traceless elements of $\Pi_n$ \cite{PZ2}.

In further prospect, the gradings are to be used e.g. for
constructing of grading preserving contractions ({\it graded
contractions}) of semisimple ${\mathcal L}$ \cite{1,2}. For
gradings involving a decomposition into a small number of grading
subspaces, this is a relatively easy task
\cite{MPT,TT1,TT2}. However for fine gradings of
algebras with ranks $\geq3$, the system of quadratic
equations for contraction parameters one needs to solve, often
gets quite large. The task of solving of such a system, would be
simplified by the knowledge of its symmetries. Symmetries
that are available are provided by those elements of
Aut$\,{\mathcal L}$ which leave the given grading invariant.

The main goal of this paper is to demonstrate that
the decomposition of $sl(n,\mathbb{C})$, as the fine grading by
$\Pi_n$, is preserved by the finite group $SL(2,\mathbb{Z}_n)$,
acting through its $n$-dimensional representation. Here
$\mathbb{Z}_n$ is the cyclic group of order $n$. We also
illustrate an application of this fact.

Special role of matrices $\Pi_n$ has been
recognized in the physical literature for a long time \cite{W}. In
more recent years there was a number of papers where the matrices
were used as a basic part of the formalism  in the development of
quantum mechanics in discrete spaces $\mathbb{Z}_n$.
The finite group $\Pi_n$ plays here the role of the discrete
Weyl group acting in an $n$--dimensional complex Hilbert space.
See \cite{ST,T1} and references quoted there.

The $\Pi_n$-grading of $sl(n,\mathbb{C}$ has other special
properties. Let us name just two:
\begin{enumerate}
\item All generators are in the same conjugacy   class of
$SL(n,\mathbb{C})$. Considered as group elements, they are of
order $n$, belonging to the Costant conjugacy class of finite
order elements, specified as $[1,1,\dots ,1]$ in the notation
introduces in \cite{K}
\item The $\Pi_n$-grading makes explicit the decompostion of
$sl(n,\mathbb{C})$ into the sum of $n+1$ Cartan subalgebra.
Indeed, if the element of ${\mathcal P}_n$ defined by a couple
$(a,b)$ ($a,b$ considered  $\mod n$), belongs to one such Cartan
subalgebra, that subalgebra is then generated by the $n-1$
elements carrying labels $(a,b)$, $(2a,2b),\dots ,$
$((n-1)a,(n-1)b)$. Clearly such elements commute and have a
non-zero determinant.
\end{enumerate}

In Section 2, the role of automorphisms of the Lie algebra in its
gradings is recalled. Grading groups of $sl(n,\mathbb{C})$ which
do not involve outer automorphisms are described in Section 3. Our
main result is in Section 4, namely the normalizers of the grading
groups. Section 5 contains an application to $sl(3,\mathbb{C})$:
It is shown that the set of 48 quadratic equations for contraction
parameters splits into just two orbits of the normalizer of the
grading group $\Pi_3$.

\section{Gradings and automorphisms of Lie algebras}
A grading of Lie algebra ${\mathcal L}$ is a decomposition of ${\mathcal
L}$
into direct sum of subspaces
\begin{equation} \label{Gr}
\Gamma\,:\quad {\mathcal L}=\bigoplus_{i\in I}{\mathcal L}_i
\end{equation}
such  that for any pair of indices $i,j\in I$ there exists  a index $k\in
I$
with the property
$$
 [{\mathcal L}_{i},{\mathcal L}_{j}]:=\{[X,Y]\ |\ X\in {\mathcal L}_{i},
Y\in {\mathcal L}_{j} \}  \subseteq {\mathcal L}_{k}
$$

\noindent A grading which cannot be further refined is called {\it fine}.

\bigskip

Gradings can be obtained by looking at Aut~${\mathcal{L}}$,
the group of all automorphisms of ${\mathcal L}$.
It consists of all non--singular linear transformations
$\phi$ of ${\mathcal L}$ as linear space
($\phi \in \mbox{GL} ({\mathcal L})$) which preserve the binary
operation in ${\mathcal L}$:
$$ \phi \, [X,Y] = [\phi \, X, \phi \, Y].$$
If $\phi$ is {\it diagonable}
 and $X$, $Y$ are its
eigenvectors with non--zero eigenvalues $\lambda$, $\mu$,
$$  \phi \, X = \lambda \, X, \quad
    \phi \, Y = \mu \, Y,  $$
then clearly
\begin{equation} \label{01}
\phi \, [X,Y] =  [\phi \, X, \phi \, Y]=
  \lambda \mu \, [X,Y].
\end{equation}
This means that the element $[X,Y]$ is either an
eigenvector of $\phi$ with eigenvalue $\lambda \mu$,
or is the zero element. The given automorphism $\phi$
thus leads to a decomposition of the linear space
${\mathcal L}$ into eigenspaces of $\phi$,
$$
{\mathcal L} = \bigoplus_{i\in I} \mbox{Ker}
               (\phi - \lambda_i \, {\rm id}),
$$
which, according to (\ref{01}), satisfies the definition of a
grading.

Refinements of a given grading, i.e. further
decompositions of the subspaces, can be obtained
by adjoining further automorphisms commuting with
$\phi$. Hence, in general, sets $\phi_1$,...,   $\phi_m$
of mutually commuting automorphisms determine
gradings.

Conversely, if a grading (\ref{Gr}) of a simple Lie algebra
${\mathcal L}$  is given, it defines
a particular Abelian subgroup Diag $\Gamma \subset
\rm{Aut} {\mathcal L}$ consisting of those automorphisms
$\phi \in GL({\mathcal L})$ which \\
(i) preserve $\Gamma, \q \phi({\mathcal L}_i) = {\mathcal L}_i$,
 \\
(ii) are diagonal,
$ \phi X = \lambda_i X  \q  \forall X \in {\mathcal L}_i,
\, i \in I,$
where $\lambda_i \neq 0$ depends only on $\phi$ and
$i \in I$.

In \cite{PZ1} an important theorem was proved:
\begin{quote} {\bf Theorem 1.}
{\it Let ${\mathcal L}$ be a finite--dimensional simple Lie
algebra over an algebraically closed field of characteristic
zero. Then the grading  $\Gamma $
is fine, if and only if the diagonal subgroup $Diag \,
\Gamma$ is a maximal Abelian group of diagonable
automorphisms (shortly MAD--group).
}
\end{quote}

A general algorithm to construct all MAD--groups
for the class of simple classical Lie algebras
over complex numbers
was given in \cite{HPP1,HPP2,HPP3}.
\footnote{Further results concerning the real forms
can be found in \cite{HPP4,HPP6}.}
These Lie algebras are Lie subalgebras of $gl(n,\mathbb{C})$,
hence their MAD--groups can be determined from
the MAD--groups of $gl(n,\mathbb{C})$ by imposing certain conditions.

The automorphisms of $gl(n,\mathbb{C})$ can be easily written
as combinations of inner and outer automorphisms.
For all $X \in gl(n,\mathbb{C})$,
\begin{quote}
 {\it inner automorphisms} have the general form
$$ \mbox{Ad}_{A} \, X = A^{-1}XA \q
    \mbox{for any} \q  A \in GL(n,\mathbb{C});$$
 {\it outer automorphims} have the general form
$$  \mbox{Out}_{C} \, X = - (C^{-1}XC)^{T} =
          \mbox{Out}_{I} \mbox{Ad}_{C} X,\q
   \mbox{where} \q  C \in GL(n,\mathbb{C}).$$
\end{quote}
Relevant properties of inner and outer automorphisms of
$gl(n,\mathbb{C})$ are summarized in the following lemma \cite{HPP1}
which allows to express MAD--groups in Aut $gl(n,\mathbb{C})$
in terms of special elements of $GL(n,\mathbb{C})$:
\begin{quote}
{\bf Lemma 2.} {\it Let $A, B, C \in GL(n,\mathbb{C})$.
\begin{enumerate}
\item Ad$_{A}$ is diagonable automorphism if and
only if the corresponding matrix $A$ is diagonable.
\item Inner automorphisms commute,
 Ad$_{A}$ Ad$_{B}$ =Ad$_{B}$ Ad$_{A}$,
if and only if there exists $q \in \mathbb{C}$ such that
\begin{equation} \label{02}
AB = q BA, \quad \text{where} \, q \, \mbox{satisfies}
\quad q^{n} = 1.
\end{equation}
\item  Out$_{C}$ is diagonable if and only if
    $C(C^{T})^{-1}$ is diagonable.
\item  Inner and outer automorphisms commute,
 Ad$_{A}$ Out$_{C}$ =Out$_{C}$ Ad$_{A}$, if and only if
$ACA^{T} = r C$; since  Ad$_{\alpha A}$  = Ad$_{A}$ for
$\alpha \neq 0$, number $r$ can be normalized to unity.
\end{enumerate}  }
\end{quote}
{\small {\bf Remark 3.}
The sets of complex $n \times n$ matrices
satisfying (\ref{02}) were to our knowledge first
studied by H. Weyl \cite{W}.

\section{MAD--groups without outer automorphisms}
In this contribution we are going to look at
the MAD--groups in Aut $gl(n,\mathbb{C})$
without outer automorphisms, i.e.
generated by inner automorphisms  (the Ad--action
in $GL(n,\mathbb{C})$ only).
It is shown in \cite{HPP1} that there exists a one-to-one correspondence
between MAD-groups without outer automorphisms and Ad-groups (to be
defined below) in $ GL(n,\mathbb{C})$.
\begin{quote}
{\bf Definition 4.} {\it A subgroup of diagonable
matrices $G \subset GL(n,\mathbb{C})$ will be called an
{\bf Ad--group} if
\begin{enumerate}
\item  for any pair $A,B \in G$ the commutator
$q(A,B) = ABA^{-1}B^{-1}$ lies in the centre
$Z = \{\alpha I_{n} \vert \alpha \in \mathbb{C}^{*} \}
\subset GL(n,\mathbb{C})$;
\item  $G$ is maximal, i.e. for each $M \notin G$
there exists $A \in G $ such that $q(A,M) \notin Z$.
\end{enumerate}}
\end{quote}

In order to describe Ad--groups we introduce
the following notation. The subgroup of $GL(n,\mathbb{C})$
containing all regular diagonal matrices will be
denoted by $D(n)$. We also define special
$k \times k$ diagonal matrices
\footnote{For $k=1$ we set $Q_1 = P_1 = 1$.}
$$Q_{k}=
\mbox{diag}(1,\omega_k,\omega_{k}^2,\ldots,
\omega_{k}^{k-1}),$$
where $\omega_{k}$ is the primitive $k$--th root of unity,
$\omega_{k} = \exp (2 \pi i/k)$, and
$$P_{k}=
\left(
\begin{smallmatrix}
0&1&0& \cdots &0&0\\
0&0&1& \cdots &0&0\\
\vdots&&&\ddots & &\\
0&0&0&\cdots& 0&1\\
1&0&0&\cdots &0&0\\
\end{smallmatrix}
\right). $$ The unitary matrices $P_{k}$, $Q_{k}$ appear in the
finite--dimensional quantum mechanics (FDQM) \cite{W,ST,T1,BI},
where their integral powers play the role of exponentiated
operators of position and momentum in the position representation.
The matrices $P_{k},$ $Q_{k}$ satisfy the identity
 (\ref{02}) with $q = \omega_{k}$,
\begin{equation} \label{Heis}
P_{k}Q_{k}=  \omega_{k} Q_{k}P_{k},
\end{equation}
which in FDQM replaces the usual Heisenberg
commutation relations.
The discrete subgroup of $GL(k,\mathbb{C})$ generated by powers of
$P_{k},$ $Q_{k}$ is the
{\it discrete Weyl (or Heisenberg) group}
of FDQM in $k$--dimensional Hilbert space ${\mathcal H}_k$.
In \cite{PZ2} this group was called the {\bf Pauli group};
it consists of $k^3$ elements
\begin{equation} \label{Pauli}
\Pi_{k} = \{ \omega_{k}^{l} Q_{k}^{i} P_{k}^{j} \vert
                     i,j, l = 0,1,\ldots,k-1 \}.
\end{equation}

The classification of Ad--groups in $GL(n,\mathbb{C})$ is given
by the following theorem \cite{HPP2}
\begin{quote}
{\bf Theorem 5.} {\it
$G \subset GL(n,\mathbb{C})$ is an Ad--group if and only if
$G$ is conjugated to one of the finite groups
$$\Pi_{\pi_1} \otimes \cdots \otimes
     \Pi_{\pi_s} \otimes D(n/\pi_{1} \ldots \pi_{s}), $$
where  $\pi_{1},\ldots,\pi_{s}$ are powers of primes and
their product $\pi_{1} \ldots \pi_{s}$ divides $n$,
with the exception of the case
$     \Pi_{2} \otimes \cdots \otimes
     \Pi_{2} \otimes D(1).$ \footnote{In this case there is
 an outer automorphism in the MAD--group.}
}
\end{quote}

The simplest form of an Ad-group is $G= D(n)$. The MAD-group corresponding
to this Ad-group gives the Cartan decomposition of $gl(n,\mathbb{C})$.

In this article we shall focus on the other extremal case, namely on the Ad-group
$$ \Pi_n\otimes D(1) = \Pi_n.$$
The corresponding fine grading decomposes $gl(n,\mathbb{C})$ into a sum of
$n^2$one--dimensional subspaces \cite{PZ1,PZ2}
\begin{equation} \label{Grgl}
\Gamma_\Pi  \, : \q gl(n,\mathbb{C}) = \bigoplus_{(r,s) \in
      \mathbb{Z}_{n} \times \mathbb{Z}_{n} }{\mathcal L}_{rs},
\end{equation}
where ${\mathcal L}_{rs}= \mathbb{C} X_{rs}$ with  $X_{rs}$ being
 the basis elements of $gl(n,\mathbb{C})$
representing $n^2$ cosets of $\Pi_n$ with respect to its
center $\{ \omega^{l} \vert l\in \mathbb{Z}_n \}$:
$$ X_{rs} = Q^{r} P^{s}. $$
Their commutators
\footnote{Henceforth the index $k$  in $\omega_k$, $P_k$,
$Q_k$ as well as explicit notation (mod $n$)
will be omitted.}
\begin{equation}\label{komutator}
[X_{rs}, X_{r' s'}] = Q^{r} P^{s} Q^{r'} P^{s'} -
     Q^{r'} P^{s'} Q^{r} P^{s} =(\omega^{sr'}- \omega^{rs'})
      X_{r+r',s + s'}
\end{equation}
clearly satisfy the grading property with the index set $I$
being the Abelian group $ \mathbb{Z}_{n} \times \mathbb{Z}_{n}$. The
binary
correspondence $((r,s),(r',s')) \mapsto (r+r',s + s')$,
$ 0\neq [{\mathcal L}_{rs},{\mathcal L}_{r's'}]
\subseteq {\mathcal L}_{r+r',s + s'} $ is the group
 multiplication in $ \mathbb{Z}_{n} \times \mathbb{Z}_{n}$
written additively modulo $n$.

The corresponding grading of $sl(n,\mathbb{C})$ contains $n^{2}-1$
subspaces
$$
sl(n,\mathbb{C}) = \bigoplus_{(r,s)\neq (0,0) }{\mathcal L}_{rs},
$$
since $Tr(X_{rs}) = 0$ except $r=s=0$.

\section{Symmetries of the fine gradings $\Gamma_\Pi$}
In this section we are going to study the symmetries of
the fine gradings (\ref{Grgl}) of $gl(n,\mathbb{C})$.
From the previous section we know
that they are induced by the Pauli group
$\Pi_n \subset GL(n,\mathbb{C})$.

Generally, the symmetry group or the automorphism group
Aut~$\Gamma \subset $ Aut ${\mathcal L}$ of the grading
(\ref{Gr}) consists of those automorphisms $\phi$ of
${\mathcal L}$ which permute the components of (\ref{Gr}),
$$ \phi {\mathcal L}_{i} = {\mathcal L}_{\overline{\phi}(i)}.$$
Here $\overline{\phi}\, : \, I \rightarrow I$ is a
permutation
of the elements of $I$, so we have a permutation
representation $\Delta_\Gamma$ of Aut~$\Gamma$,
$$ \bar{\phi} = \Delta_{\Gamma}(\phi), \q
\phi \in \text{Aut}\ \Gamma. $$
The kernel of $\Delta_{\Gamma}$ is the stabilizer of
$\Gamma$ in Aut $\Gamma$,
$$ \text{Stab}\ \Gamma = \ker \Gamma =
\{ \phi \in \text{Aut}\ {\mathcal L}\ \vert \ \phi {\mathcal L}_{i}=
  {\mathcal L}_{i}\, \; \forall i \in I \}.
$$
It is a normal subgroup of Aut $\Gamma$ with quotient group
isomorphic to the group of permutations of $I$,
$$ \text{Aut}\ \Gamma / \text{Stab}\ \Gamma \simeq
  \Delta_\Gamma \ \text{Aut}\ \Gamma.
$$
For fine gradings Stab $\Gamma = {\mathcal G}$.
The {\it symmetry group} Aut $\Gamma$ is by definition
\cite{PZ1} the normalizer of ${\mathcal G}$ in Aut $gl(n,\mathbb{C})$:
$$ {\mathcal N}({\mathcal G}) = \text{Aut}\ \Gamma =
 \{ \phi \in \text{Aut}\ gl(n,\mathbb{C})\ \vert\ \phi {\mathcal G}
  \phi^{-1} \subset {\mathcal G} \}.
$$

Why do we look for the symmetry group Aut $\Gamma$ ?
We know that
\begin{quote}
(1) elements of Aut $\Gamma$ / Stab $\Gamma$ permute
the grading subspaces,\\
(2) given a grading subspace, the action of
Aut $\Gamma$ / Stab $\Gamma$ will yield some other
grading subspaces.
\end{quote}
So its knowledge may give us the way to construct the
grading decomposition (\ref{Gr}) from one or a small
number of starting subspaces. Aut $\Gamma$ / Stab $\Gamma$
may also be valuable as a symmetry of the contraction
equations which enables to lower their number and so
simplify their solution.

Let us denote the  MAD--group  $ \text{Ad}_{\Pi_n}$
by  ${\mathcal P}_n $. It is an Abelian subgroup
of Aut~$gl(n,\mathbb{C}) = GL(n^2,\mathbb{C})$ with generators Ad$_P$,
Ad$_Q$,
$${\mathcal P}_n  = \{ \text{Ad}_{Q^{i}P^{j}} \vert (i,j) \in
   \mathbb{Z}_{n} \times \mathbb{Z}_{n} \}.
$$
It is obvious that $n^2$ elements of ${\mathcal P}_n$
stabilize the grading: namely, taking the generators
Ad$_P$, Ad$_Q$, one has
$$ \text{Ad}_{P} \, X_{rs} = P Q^{r} P^{s} P^{-1} =
    \omega^{r} X_{rs}, \q
\text{Ad}_{Q} \, X_{rs} = Q Q^{r} P^{s} Q^{-1} =
    \omega^{-s} X_{rs}
$$
and ${\mathcal P}_{n} = \text{Stab}\ \Gamma_\Pi$ since
${\mathcal P}_{n}$ is maximal.

In order to describe the quotient group
$ {\mathcal N}({\mathcal P}_n)/{\mathcal P}_n$ we note that its
elements are classes of equivalence in
$ {\mathcal N}({\mathcal P}_n)$ given by
$$ \phi \sim \psi \q \text{if and only if}
 \q \phi \psi^{-1} \in {\mathcal P}_n.$$
Let $\phi \sim \psi $, i.e. $\phi \psi^{-1}=\beta$
for some $\beta \in  {\mathcal P}_n$. Using the commutativity
of ${\mathcal P}_n$ we have
$$\phi^{-1} \alpha \phi = \psi^{-1}\beta^{-1} \alpha
  \beta \psi =\psi^{-1} \alpha \psi$$
for any  $\alpha \in  {\mathcal P}_n$. On the other hand,
let $\phi^{-1} \alpha \phi = \psi^{-1} \alpha \psi$
for any $\alpha \in  {\mathcal P}_n$. Then $\phi \psi^{-1}$
commutes with every element in ${\mathcal P}_n$ and
therefore $\phi \psi^{-1} \in {\mathcal P}_n.$
It means that
$$
\phi \sim \psi \q \text{if and only if}\q
\phi^{-1} \alpha \phi = \psi^{-1} \alpha \psi \q
\text{for any}\q \alpha \in  {\mathcal P}_n.$$
Since the group ${\mathcal P}_n$ has only two generators
Ad$_P$ and Ad$_Q$, the previous condition can be
rewritten
\begin{equation} \label{equiv}
\phi \sim \psi \q \text{if and only if} \q
\phi^{-1} \text{Ad}_{P} \phi = \psi^{-1} \text{Ad}_{P} \psi
\q \text{and}\q \phi^{-1} \text{Ad}_{Q} \phi = \psi^{-1} \text{Ad}_{Q}
\psi.
\end{equation}

If $\phi$ belongs to the normalizer $ {\mathcal N}({\mathcal P}_n)$,
then there exist elements $a,b,c,d$ in the cyclic group
$\mathbb{Z}_n$ such that
$$
\phi^{-1} \text{Ad}_{Q} \phi=\text{Ad}_{Q^{a}P^{b}}
\q \text{and} \q
\phi^{-1} \text{Ad}_{P} \phi=\text{Ad}_{Q^{c}P^{d}}.
$$
Thus to any equivalence class a quadruple of indices
is assigned. Denote this assignment by $\Phi$.
According to (\ref{equiv}), quadruples assigned to distinct
classes are different. We shall see that it is convenient to
write the quadruple as a matrix
 $$ \Phi(\phi) =
\left(
\begin{array}[c]{cc}
a&b\\
c&d\\
\end{array}
\right)
\q \text{with entries from} \q  \mathbb{Z}_n.
$$
Suppose that the equivalence classes containing the
automorphisms $\phi_1$ and $\phi_2$ correspond to the
quadruples $a_1,b_1,c_1,d_1$ and $a_2,b_2,c_2,d_2$,
respectively. Computing the quadruple assigned to
the composition $\phi_1 \phi_2$
$$
(\phi_1 \phi_2)^{-1}\text{Ad}_{Q}(\phi_1 \phi_2) =
(\phi_{2}^{-1}\text{Ad}_{Q} \phi_2)^{a_1}
(\phi_{2}^{-1}\text{Ad}_{P} \phi_2)^{b_1}= $$
$$
(\text{Ad}_{Q^{a_2}P^{b_2}})^{a_1}
(\text{Ad}_{Q^{c_2}P^{d_2}})^{b_1}=
\text{Ad}_{Q^{a_{2}a_1}P^{b_{2}a_1}}
\text{Ad}_{Q^{c_{2}b_1}P^{d_{2}b_1}}
 =\text{Ad}_{Q^{a_{1}a_{2}+b_{1}c_{2}}
  P^{a_{1}b_{2}+b_{1}d_{2}}},
$$
$$(\phi_1 \phi_2)^{-1}\text{Ad}_{P}(\phi_1 \phi_2) =
\text{Ad}_{Q^{a_{1}a_{2}+b_{1}c_{2}}
  P^{a_{1}b_{2}+b_{1}d_{2}}}
$$
we see that to the automorphism $\phi_1 \phi_2$
the product matrix is assigned,
$$ \Phi(\phi_1 \phi_2) =\Phi(\phi_1)\Phi(\phi_2).
$$
Thus $\Phi$ is an injective homomorphism of the quotient
group
$ {\mathcal N}({\mathcal P}_n)/{\mathcal P}_n$.

Let $\phi\in {\mathcal N}({\mathcal P}_n)$ be an {\it inner} automorphism,
say   Ad$_A$ with the corresponding
matrix
 $$ \Phi(\phi) =
\left(
\begin{array}[c]{cc}
a&b\\
c&d\\
\end{array}
\right)
\q \text{with entries from}\ \mathbb{Z}_n.
$$
Then
\begin{equation} \label{tr3}
\text{Ad}_{A}^{-1}\text{Ad}_{Q}\text{Ad}_{A} =
\text{Ad}_{A^{-1}QA}=\text{Ad}_{Q^{a}P^{b}}
\q \text{implies} \q A^{-1}QA = \mu Q^{a}P^{b},
\end{equation}
\begin{equation} \label{tr4}
\text{Ad}_{A}^{-1}\text{Ad}_{P}\text{Ad}_{A} =
\text{Ad}_{A^{-1}PA}=\text{Ad}_{Q^{c}P^{d}}
\q \text{implies} \q A^{-1}PA = \nu Q^{c}P^{d}
\end{equation}
for some $\mu, \nu \in \mathbb{C}^*$.
Multiplying the
equations (\ref{tr3}) and (\ref{tr4}) by $PA$ and
$QA$ from the right, respectively, and using
the relation $PQ = \omega QP$, we obtain
$$PQA=\mu\nu AQ^cP^dQ^aP^b=
\omega^{ad}\mu\nu Q^{a+c}P^{b+d}$$
and
 $$QPA = \mu \nu AQ^aP^bQ^cP^d=
\omega^{bc}\mu\nu Q^{a+c}P^{b+d}$$
Since $PQA=\omega QPA$ we obtain the identity
$$
\omega^{ad-1} = \omega^{bc}, \q \text{i.e.} \q
ad - 1 = bc \ (mod\ n),$$
hence $$
\det \Phi(\phi)=1.
$$
A simple computation further shows that for this inner
 automorphism one has
$$ A^{-1} X_{rs}A = \rho X_{r's'},$$ where $|\rho |=1$ and
\begin{equation} \label{tr5}
(r',s')= (r,s)
\left(
\begin{array}[c]{cc}
a&b\\
c&d\\
\end{array}
\right).
\end{equation}

Consider now the outer automorphism $Out_I\ X:= -X^T$.
Because of
$$
(\text{Out}_{I})^{-1} \text{Ad}_{Q}\text{Out}_{I}=
\text{Ad}_{(Q^{-1})^{T}} =\text{Ad}_{Q^{-1}} \in {\mathcal P}_n,
$$
$$
(\text{Out}_{I})^{-1} \text{Ad}_{P}\text{Out}_{I}=
\text{Ad}_{(P^{-1})^{T}} =\text{Ad}_{P}  \in {\mathcal P}_n,
$$
the automorphism $Out_I$  belongs to the normalizer
$ {\mathcal N}({\mathcal P}_n)$.
The matrix corresponding to $Out_I$ is
 $$ \Phi(\text{Out}_{I}) =
\left(
\begin{array}[c]{cc}
-1&0\\
0&1\\
\end{array}
\right)
 \q \text{and} \q
\det \Phi(\text{Out}_{I}) = -1.
$$
Note that
$$\text{Out}_{I}\ X_{rs} = - \omega^{-rs}\ X_{r,-s}.$$
corresponds to the permutation of indices
$$ (r,s) \mapsto (r,-s).$$
Any other outer automorphism $\phi$ from  $ {\mathcal N}({\mathcal P}_n)$
is the composition of $\text{Out}_{I}$  and an inner
automorphism from  $ {\mathcal N}({\mathcal P}_n)$ and thus
$\det \Phi(\phi) = -1$.
We conclude with

\bigskip

\begin{quote}
{\bf Proposition 6.}\ \ \ {\it $\Phi$  is an injective homomorphism of
${\mathcal N}({\mathcal P}_n)/{\mathcal P}_n$
  into the group
$$H= \left\{\left( \begin{array}{cc}
               a&b\\
               c&d
              \end{array}
\right)\ |\  a,b,c,d\in\mathbb{Z}_n,\    ad-bc=\pm 1  \
{\rm mod}\ n, \right\}.
$$}
\end{quote}

\bigskip

The group $H$ contains as its subgroup the group of matrices with determinant $+1$.
This group is usually denoted by $SL(2,\mathbb{Z}_n)$.
Note that
$\mathbb{Z}_n$ is a field iff $n$ is prime.
Clearly
 $$H= SL(2,\mathbb{Z}_n)\  \bigcup\
\left( \begin{smallmatrix}
               1&0\\
               0&-1
              \end{smallmatrix}
\right)\ SL(2,\mathbb{Z}_n).$$

Let us briefly show  that for any $n\in \mathbb{N}$, the group
$SL(2,\mathbb{Z}_n)$ is generated by two matrices
$$A=\left( \begin{array}{rr}
               1&0\\
               1&1
              \end{array}
\right)\quad {\rm and} \quad B=\left( \begin{array}{rr}
               0&-1\\
               1&0
              \end{array}
\right).$$
In order to show it, we have to realize that in the ring $\mathbb{Z}_n$,
the matrices  $A$ and $B$ are of orders $n$ and $4$, respectively.
The matrix  $C=A^T$ is generated by $A$ and $B$:
$$C=\left( \begin{array}{rr}
               1&1\\
               0&1
              \end{array}
\right)= \left( \begin{array}{rr}
               0&-1\\
               1&0
              \end{array}
\right)\left( \begin{array}{rr}
               1&0\\
               1&1
              \end{array}
\right)^{n-1}\left( \begin{array}{rr}
               0&-1\\
               1&0
              \end{array}
\right)^3 .$$
Moreover,  any matrix
of  $SL(2,\mathbb{Z}_n)$ satisfies
\begin{equation}\label{euclid}
 \left( \begin{array}{rr}
               a&b\\
               c&d
              \end{array}
\right)=\left( \begin{array}{rr}
               1&0\\
               1&1
              \end{array}
\right)\left( \begin{array}{cc}
               a&b\\
               c-a&d-b
              \end{array}
\right)=\left( \begin{array}{cc}
               1&1\\
               0&1
              \end{array}
\right)\left( \begin{array}{cc}
               a-c&b-d\\
               c&d
              \end{array}
\right).
\end{equation}\
Now recall that Euclid's algorithm for finding the greatest
common divisor of integers is based on the trivial fact
that $gcd(x,y)=gcd(x-y,y)$. By several repetitions of this
rule, where we replace the pair of non--negative integers
$\{x,y\}, x\geq y,$ by another pair of non--negative
integers $\{x-y,y\}$, the Euclid's algorithm  gives
finally a pair of integers, where one of them is 0 and
the other is $gcd(x,y)$.

Denote $gcd(a,c)=s$. Then by suitable applications  of (\ref{euclid}) we obtain
$$\left( \begin{array}{cc}
               a&b\\
               c&d
              \end{array}
\right)= A^{k_1}C^{l_1}\dots A^{k_p}C^{l_p} T$$
where $k_1,l_1,\dots ,k_p,l_p\in \mathbb{N}_0$ and $T$ is
a matrix with $\det\ T=1$, of the form
$$T= \left( \begin{array}{cc}
               s&t\\
               0&u
              \end{array}
\right)\quad{\rm or}\quad T=
\left( \begin{array}{cc}
               0&v\\
               s&w
              \end{array}
\right).$$
But any  matrix $T$ of such form is a product of several matrices $A$,$B$,
and $C$, since
$$\left( \begin{array}{cc}
               s&t\\
               0&u
              \end{array}
\right)= \left( \begin{array}{cc}
               1&1\\
               0&1
              \end{array}\right)^{s(t-1)}
\left( \begin{array}{cc}
               1&0\\
               1&1
              \end{array}\right)^u
\left( \begin{array}{rc}
               0&1\\
              -1&0
              \end{array}\right)
\left( \begin{array}{cc}
               1&0\\
               1&1
              \end{array}\right)^s$$
and
 $$\left( \begin{array}{cc}
               0&v\\
               s&w
              \end{array}
\right)= \left( \begin{array}{rc}
               0&1\\
              -1&0
              \end{array}\right)
\left( \begin{array}{rr}
               -s&-w\\
               0&v
              \end{array}\right).
$$

This proves that $SL(2,\mathbb{Z}_n)$ has two generators, the matrices $A$
and $B$.

\medskip

Now we shall present two  elements of the
normalizer --- special unitary
$n\times n$-matrices inducing Ad-actions which represent
generating elements of $SL(2,\mathbb{Z}_n)$.

{\bf Example 7.}\ \  Since the matrices $Q$ and $P$
have the same spectra, they are similar with a similarity matrix $S$
 such that $S^{-1}PS=Q$.
Such $S$ is not determined uniquely. We choose for the
matrix $S$ the Sylvester matrix defined as follows
 $$S_{ij} = \omega^{-ij}, \ \ \ {\rm for}\ \ i,j\in \mathbb{Z}_n.$$
It is easy to verify that $S^2$ is a parity operator
 $$S^2_{ij}= \delta_{i,-j}\ \ {\rm for}\ \ i,j\in \mathbb{Z}_n \ \
 \text{such that}\ \ S^4=I.$$
 Note that the indices (and operations on them) are always
 considered to be elements of the ring $\mathbb{Z}_n$.
Let us verify that $Ad_S$ belongs to the normalizer.
 Note that $S$ is a symmetric matrix
 and therefore $Q=Q^T = (S^{-1}PS)^T= SP^TS^{-1}=
SP^{-1}S^{-1}$, which implies
 $S^{-1}QS=P^{-1}$. Now  we can easily check the conditions
 on $Ad_S$ to be in the normalizer:
$$ (Ad_S)^{-1}Ad_QAd_S = Ad_{S^{-1}QS}=Ad_{P^{-1}} \in
{\mathcal P}_n$$  and
 $$ (Ad_S)^{-1}Ad_PAd_S = Ad_{S^{-1}PS}=Ad_Q \in
{\mathcal P}_n$$
  By (\ref{tr3}) and (\ref{tr4}) the matrix corresponding to
 $Ad_S$ is
 $$
\Phi(Ad_S)=\left(\begin{array}{cr}0&-1\\1&0\end{array}\right)
$$

{\bf Example 8.}  For this example we shall use the
similarity of matrices $P$ and $PQ$. Put $\varepsilon =1$,
 if $n$ is odd, and $\varepsilon=\sqrt{\omega}$, if $n$ is even.
 Denote by
 $$
D=\text{diag}(d_0,d_1,\dots, d_{n-1}), \ \ {\rm where}\
 d_j=\varepsilon^{-j}\omega^{-(^{j}_{2})},\ \ {\rm for}\ \
j\in \mathbb{Z}_n
$$
 It is easy to see that
 $Q=D^{-1}QD$ and $PQ=\varepsilon D^{-1}PD $; it implies
 $$
(Ad_D)^{-1}Ad_QAd_D = Ad_{D^{-1}QD}= Ad_Q \in
{\mathcal P}_n
$$  and
 $$ (Ad_D)^{-1}Ad_PAd_D = Ad_{D^{-1}PD}=Ad_{PQ} \in
 {\mathcal P}_n$$ which means that $Ad_D$ belongs to the
 normalizer.    By (\ref{tr3}) and (\ref{tr4})
the matrix assigned to $Ad_D$ is therefore
 $$\Phi(Ad_D) =
 \left(\begin{array}{cc}1&0\\1&1\end{array}\right)$$

\medskip
The homomorphism $\Phi$ maps three elements of the normalizer
$Out_I$, $Ad_S$, and $Ad_D$ into three matrices generating the whole group
$H$. This observation together with the Proposition 6  gives us\\[1mm]

\begin{quote}
{\bf Theorem 10.} \ \ \ {\it The factor group
${\mathcal N}({\mathcal P}_n)/{\mathcal P}_n$ is isomorphic to the group
$$ \left\{\left( \begin{array}{cc}
               a&b\\
               c&d
              \end{array}
\right)\ |\ a,b,c,d \in \mathbb{Z}_n,\ ad-bc=\pm 1 \ {\rm mod}\ n\right\}=
SL(2,\mathbb{Z}_n)\  \bigcup\
\left( \begin{smallmatrix}
               1&0\\
               0&-1
              \end{smallmatrix}
\right)\ SL(2,\mathbb{Z}_n).
$$}
\end{quote}

 The direct consequence of Theorem 10 is \\[1mm]

\begin{quote}
{\bf Corollary  11.} \ \ \ {\it The normalizer
${\mathcal N}({\mathcal P}_n)$
of the group ${\mathcal P}_n$ is generated by \\
 \centerline{$Out_I,Ad_S, Ad_D, Ad_Q  \ \  and\ \
 Ad_P$}}
\end{quote}

 \bigskip

If $n$ is prime, i.e. $\mathbb{Z}_n$ is a field, we can
use the Bruhat decomposition of $SL(2,\mathbb{Z}_n)$ and
 explicitly  describe the normalizer. It enables us to count the number of its elements.
The group $SL(2,\mathbb{Z}_n)$ is the union of two disjoint sets
$$\left\{\left(\begin{array}{cc}1&0\\a&1\end{array}\right)
  \left(\begin{array}{cc}b&0\\0&b^{-1}\end{array}\right)\ |\
  a\in \mathbb{Z}_n, b\in \mathbb{Z}_{n}^{*}\right\}$$
  and
 $$\left\{\left(\begin{array}{cc}1&0\\a&1\end{array}\right)
  \left(\begin{array}{cc}b&0\\0&b^{-1}\end{array}\right)
  \left(\begin{array}{cr}0&-1\\1&0\end{array}\right)
 \left(\begin{array}{cc}1&0\\c&1\end{array}\right)\ |\
  a,c\in \mathbb{Z}_n, b\in \mathbb{Z}_{n}^{*}\right\}.$$

\bigskip

\begin{quote}
{\bf Corollary 12.} {\it \ \ Let $n$ be a prime.
Any element of the normalizer
${\mathcal N}({\mathcal P}_n)$ has the form

\medskip

\centerline{$Ad_A$ or $Out_IAd_A$, where
$A=D^iM^jQ^kP^l$ or
$A=D^iM^jSD^sQ^kP^l$}

\medskip

and $M$ is a permutation matrix described in the Appendix.
\\ The normalizer   ${\mathcal N}({\mathcal P}_n)$
 has therefore $2(n^2-1)n$ elements for an odd prime $n$ 
and 24 elements for $n=2$ (here the outer automorphism 
does not play any role)}
\end{quote}

\bigskip

Summarizing, for any natural number $n$ we described the
 generators of the normalizer 
${\mathcal N}({\mathcal P}_n)/{\mathcal P}_n$. 
For $n$ prime we moreover were able to determine 
the cardinality of normalizer, using the
Bruhat decomposition of the group  $ SL(2, \mathbb{Z}_n)$.
For the explicit description of the special
$n$--dimensional representation of $SL(2, \mathbb{Z}_n)$,
where $n$ is  prime of the form $n=4K \pm 1$, see
\cite{BI}.

\section{Graded contractions of $sl(3,\mathbb{C})$}
In this section we want to illustrate how the explicit knowledge
of the normalizer of a fine grading can simplify and, indeed,
bring further insight into the structure of the problem of finding
all graded contractions of $sl(3,\mathbb{C})$. Practically one
needs to solve a system of 48 quadratic equations involving 28
contraction parameters. The normalizer is the symmetry group of
that system. It turns out that there are only two orbits of the
normalizer among the 48 quadratic equations. For more conventional
approach to this problem see \cite{W1,W2}.
\medskip

Let ${\mathcal L}=\bigoplus_{i\in I} {\mathcal L}_{i}$ be a
grading decomposition of a Lie algebra with the commutator
$[\,\cdot\,,\cdot\,]$. Definition of the contracted commutator of
the algebra involves the contraction parameters $\varepsilon_{ij}$
for $i,j \in I$ and the old commutator. The new bilinear mapping
of the form $$ [x,y]_{new} := \varepsilon_{ij} [x,y] \qquad {\rm
for\ all\ } x\in {\mathcal L}_{i}, \ y\in {\mathcal L}_{j} $$ is a
commutator on the same vector space ${\mathcal L}$.

To satisfy antisymmetry of the commutator we have to choose
$\varepsilon_{ij}=\varepsilon_{ji}$. To satisfy the Jacobi
identity one has to solve a system of quadratic equations for the
unknown contraction parameters $\varepsilon_{ij}$.

Let us illustrate this problem on the graded algebra
$sl(3,\mathbb{C})= \bigoplus_{(i,j) \neq (0,0)} {\mathcal
L}_{ij}$, which has 8 one-dimensional graded subspaces ${\mathcal
L}_{ij} = \mathbb{C} X_{ij}$, where $0\leq i,j\leq 2$. For
example, for the triple of vectors $X_{(0,1)}$, $X_{(0,2)}$  and
$X_{(1,0)}$ the Jacobi identity has the form $$
[X_{(0,1)},[X_{(0,2)},X_{(1,0)}]_{new}]_{new} + {\rm
 cyclically}\ = 0.
$$
The commutation relations \ref{komutator} give us
$$
\varepsilon_{(02)(10)}\varepsilon_{(01)(12)} (\omega -
 1)(\omega ^2 -1) X_{(1,0)} +
 \varepsilon_{(10)(01)}\varepsilon_{(02)(11)} (1-\omega)
 (\omega ^2 -1) X_{(1,0)} = 0
$$
 and therefore
 \begin{equation}\label{prvni}
\varepsilon_{(02)(10)}\varepsilon_{(01)(12)}
-\varepsilon_{(10)(01)}\varepsilon_{(02)(11)}=0
 \end{equation}

 For all possible triples of basis elements $X_{(i,j)}$ we have to
 write similar  equations. There are ${(^{8}_{3})} = 56$ triples.
 Since triples of the form  $X_{(a,b)}$, $X_{(c,d)}$  and
$X_{(e,f)}$, with $a+c+e=0\ (mod\ 3)$ and $b+d+f=0\ (mod\ 3)$
satisfy $[X_{(a,b)},[X_{(c,d)},X_{(e,f)}]]=0$, we have in fact
only 48 equations.

The Jacobi identity for the triple $X_{(0,1)}$, $X_{(1,0)}$  and
$X_{(1,1)}$, is the equality
\begin{equation}\label{druha}
\varepsilon_{(10)(11)}\varepsilon_{(01)(21)}
-\varepsilon_{(11)(01)}\varepsilon_{(10)(12)}=0\ .
\end{equation}
The triple of indices $(0,1)$, $(1,0)$ and $(1,1)$ is
distinguished by the property that the indices of any epsilon
appearing in \eqref{druha} are linearly independent over the
field $\mathbb{Z}_3$. Quite different are the indices  in
\eqref{prvni}. There the pair of indices $(0,1)$ and $(0,2)$ is
linearly dependent over the field $\mathbb{Z}_3$. These two cases
exhaust all distinct possibilities for the choice of triples in
the Jacobi identity.

Consider now the mappings on the index set $I$ defined by
(\ref{tr5})
  $$(i,j) \mapsto (i,j)A\qquad {\rm where}\
A\in  SL(2,\mathbb{Z}_3)$$
 Applying such a mapping with a fixed matrix $A$
 to the indices occurring in equation
  (\ref{prvni}) we obtain a new equation corresponding
to the Jacobi identity for another
triple of grading subspaces.
 If we gradually apply all 24 matrices from $SL(2,\mathbb{Z}_3)$
 to the equations (\ref{prvni}) and (\ref{druha}),
 we obtain all
 48 quadratic equations which should be satisfied.
In this way the symmetries of the system of equations
are directly seen.

\subsection*{Appendix}

Consider $n \times n$ matrices $M_s$ defined by
 $$
(M_s)_{i,j} = \delta_{i,sj}\ \ \ {\rm for}\ \ 
i,j\in Z_n,
$$
where $n$ is a prime and $s\in Z_{n}^{*}  = Z_n\backslash\{0\}$.
 Since $Z_n$ is a field, the equation $is=j$ has one
 solution $i$ for  fixed $s$ and $j$ and similarly one
 solution $j$ for fixed $s$ and $i$. It means that 
each matrix $M_s$ has exactly one 1 in each
 column and row; $M_s$ are thus  permutation matrices.
Moreover, $M_1$ is the unit matrix and $M_sM_t = M_{st}$
holds:
 $$
(M_sM_t)_{ij} = \sum_{k=0}^{n-1}(M_s)_{ik}(M_t)_{kj}=
 \sum_{k=0}^{n-1}\delta_{i,sk}\delta_{k,tj}=
\delta_{i,jst}=(M_{st})_{ij},
$$
hencet the matrices $M_s$, $s=1,2,\dots,n-1$ form 
a representation of a multiplicative Abelian group.
 This group has only one generator, because
for any finite  field $(F, +,\cdot)$ the
 group $(F^{*}=F\backslash\{0\}, \cdot)$ is cyclic. 
In particular for $Z^{*}_n$ there exists an element, 
say $a$, such that
 $Z^{*}_n=\{a^i\ | \ i=1,2,\dots, n-1\}$. 
Therefore $M_1, \dots , M_{n-1}$ is a cyclic group
of order $n-1$ with generator  $M_a$ and we have
$M_{a^s}= M_{a}^{s}$.  \\[1mm]

We shall show that $Ad_{M_s}$  belongs to  the normalizer
 ${\mathcal N}({\mathcal P}_n)$. We use the fact that $M_s$ is
a permutation matrix and therefore $M_s^{-1}=M_s^T$.
 Let us compute $M_s^{-1}QM_s$ and   $M_s^{-1}PM_s$:
 $$
(M_s^{-1}QM_s)_{ij}=
\sum_{k=0}^{n-1}(M_s^T)_{ik}(QM_s)_{kj}=
\sum_{k=0}^{n-1}
 \sum_{r=0}^{n1}(M_s)_{ki}Q_{kr}(M_s)_{rj}=
$$

 $$~~~~~~~~~~~~~~~~~~~~~~=\sum_{k=0}^{n-1}
 \sum_{r=0}^{n-1}\delta_{k,is}\omega^k\delta_{kr}\delta_{r,sj}=
\sum_{k=0}^{n-1}\delta_{k,is}\omega^k\delta_{k,sj}=
\delta_{ij}\omega^{si}= (Q^s)_{ij},
$$
hence
\begin{equation}\label{housle}
M_s^{-1}QM_s = Q^s\ .
\end{equation}
Further,
 $$
(M_s^{-1}PM_s)_{ij}=\sum_{k=0}^{n-1} (M_s^T)_{ik}(PM_s)_{kj}
 =
\sum_{k=0}^{n-1}\sum_{r=0}^{n-1}(M_s)_{ki}P_{kr}(M_s)_{rj}= $$
$$= \sum_{k=0}^{n-1}\sum_{r=0}^{n1}
 \delta_{k,is}\delta_{k,r+1} \delta_{r,sj}=
\sum_{r=0}^{n-1}\delta_{is,r+1}\delta_{r,sj}
 =\delta_{is, js+1}
$$
and noting that $\delta_{is,js+1} = \delta_{i,j+t}$, 
where $st=1$ in $Z_n$, we have
 \begin{equation}\label{kridla}
 M_s^{-1}PM_s=P^{s^{-1}}\ .
 \end{equation}
Equations (\ref{housle}) and (\ref{kridla}) mean that
all $Ad_{M_s}$ belong to the normalizer and 
that the matrix assigned to $M_s$ is
 $$
\Phi(M_s)= \left(\begin{array}{cc}s&0\\
0&s^{-1}\end{array}
 \right).
$$

\subsection*{Acknowledgements}

M.H., E.P. and J.T. gratefully acknowledge the support
 of the Ministry of Education of Czech Republic 
under the research contract MSM210000018. 
M.H. and E.P. further thank the Grant
Agency of Czech Republic for the support under the contract
201/01/0130. E.P. acknowledges the hospitality of CRM,
 Montreal and very fruitful discussion with A. Penskoi. 
J.P. was partiallyb supported by the National Science 
and Engineering Research Council of Canada.

\end{document}